\documentclass[jap,preprint,graphicx,unsortedaddress,showpacs]{revtex4}

\renewcommand{\baselinestretch}{1.7}
\usepackage{hyperref}
\usepackage[english]{babel}
\usepackage[T1]{fontenc}
\usepackage[latin1]{inputenc}
\usepackage[dvips]{graphicx}
\usepackage{mathenv}
\usepackage{array}
\usepackage{bm}
\usepackage{fancyhdr}
\usepackage{makeidx}
\usepackage{subfigure}
\usepackage{babel}
\usepackage[v2]{xy}
\usepackage{epsfig}
\usepackage{color}
\hypersetup{backref=true,pagebackref=true,hyperindex=true,colorlinks=true,breaklinks=true,urlcolor=
blue,linkcolor=blue,bookmarks=true,bookmarksopen=true}

\begin{document}
\renewcommand{\baselinestretch}{1.5}
\title{Spin-dependent diffraction at ferromagnetic/spin spiral interface}
\date{\today}
\author{A. Manchon}
\affiliation{SPINTEC, URA 2512 CEA/CNRS, CEA/Grenoble, 38054 Grenoble Cedex 9, France}
\author{N. Ryzhanova}
\affiliation{SPINTEC, URA 2512 CEA/CNRS, CEA/Grenoble, 38054 Grenoble Cedex 9, France}
\affiliation{Department of Physics, M. V. Lomonosov Moscow State University, 119899 Moscow, Russia}
\author{A. Vedyayev}
\affiliation{SPINTEC, URA 2512 CEA/CNRS, CEA/Grenoble, 38054 Grenoble Cedex 9, France}
\affiliation{Department of Physics, M. V. Lomonosov Moscow State University, 119899 Moscow, Russia}
\author{B. Dieny}
\affiliation{SPINTEC, URA 2512 CEA/CNRS, CEA/Grenoble, 38054 Grenoble Cedex 9, France}

\begin{abstract}
Spin-dependent transport is investigated in ballistic regime through the interface between a ferromagnet and a spin spiral. We show that spin-dependent interferences lead to a new type of diffraction called "spin-diffraction". It is shown that this spin-diffraction leads to local spin and electrical currents along the interface. This study also shows that in highly non homogeneous magnetic configuration (non adiabatic limit), the contribution of the diffracted electrons is crucial to describe spin transport in such structures.
\end{abstract}
\maketitle
\maketitle
\clearpage

The recent observations of current-induced domain wall motion \cite{DW} and the investigation of this phenomenon by micromagnetic simulations \cite{sim} have underlined the question of spin transport in non homogeneous systems. Although proposed very early \cite{berger78}, current-induced domain wall motion (DWM) has attracted much attention because of its great application potential but also because of the deep fundamental question of the role of electron spin motion in temporally and spacially varying magnetic structures. Zhang et al. \cite{zhang04} proposed an analytical formula for spin torque in slowly varying structures (see eq. 9 in Ref.~\onlinecite{zhang04}):
\begin{equation}\label{e:e0}
\bm{T}=a\frac{\partial\bm{M}}{\partial t}+b\bm{M}\times\frac{\partial\bm{M}}{\partial t}-c_1\bm{M}\times[\bm{M}\times(\bm{j}_e.\nabla)\bm{M}]-c_2\bm{M}\times(\bm{j}_e.\nabla)\bm{M}
\end{equation}
where ${\bf M}$ is the magnetization unit vector, $a$ is the renormalization factor of the gyromagnetic ratio and $b$ is the renormalization factor of the damping parameter in the Landau-Lifshitz-Gilbert equation. The last two terms represent the spin transfer torque in spacially varying magnetic structure. The prefactors $c_1$ and $c_2$ are respectively proportionnal to the adiabatic and non adiabatic contribution. For smooth enough non-homogeneities of the magnetic structure, the adiabatic approximation is usually assumed : the electron spin follows the local magnetization producing a small torque proportional to the spatial derivative of the magnetization \cite{xiao06,tserko}. In this case, it is usually accepted that the non adiabatic term $c_2$ is small but cannot be neglected in domain wall experiments \cite{thiaville,waintal}. Furthermore, in spacially non homogeneous systems, $c_1$ and $c_2$ are non local coefficients \cite{xiao06,tserko,tatara}: at each point of the structure, one has to consider the contribution of all the electrons flowing through the structure.\par

In this article, we propose to study the non adiabatic regime in an "academic" system. We consider two adjacent layers: the left one ($z$<0, F) with a homogenous magnetization $\bm{P}=P\bm{z}$ and the right one ($z$>0, SS - for Spin Spiral) with a 2D helical magnetization contained in the ($x$,$z$) plane $\bm{M}=M(\sin\theta(x)\bm{x}+\cos\theta(x)\bm{z})$, where $\theta(x)=\theta_0 + Qx$. The interface lies in the ($x$,$y$) plane and $z$ is perpendicular to the interface. We consider that the regions are semi-infinite and respectively connected to a ferromagnetic and a spin spiral reservoir. The bias voltage V is applied across the interface.\par

Such helical spin structures exist in some compounds such as MnSi \cite{MnSi}, oxide materials such as SrFeO$_3$, NaCuO \cite{ox}, rare-earth based compounds \cite{re} or $\gamma$-iron \cite{marsman}. This helical structure can be also a simplified picture of narrow stripe domains with domain wall width comparable to domain width. \par

\begin{figure}[h]
	\centering
		\includegraphics[width=8cm]{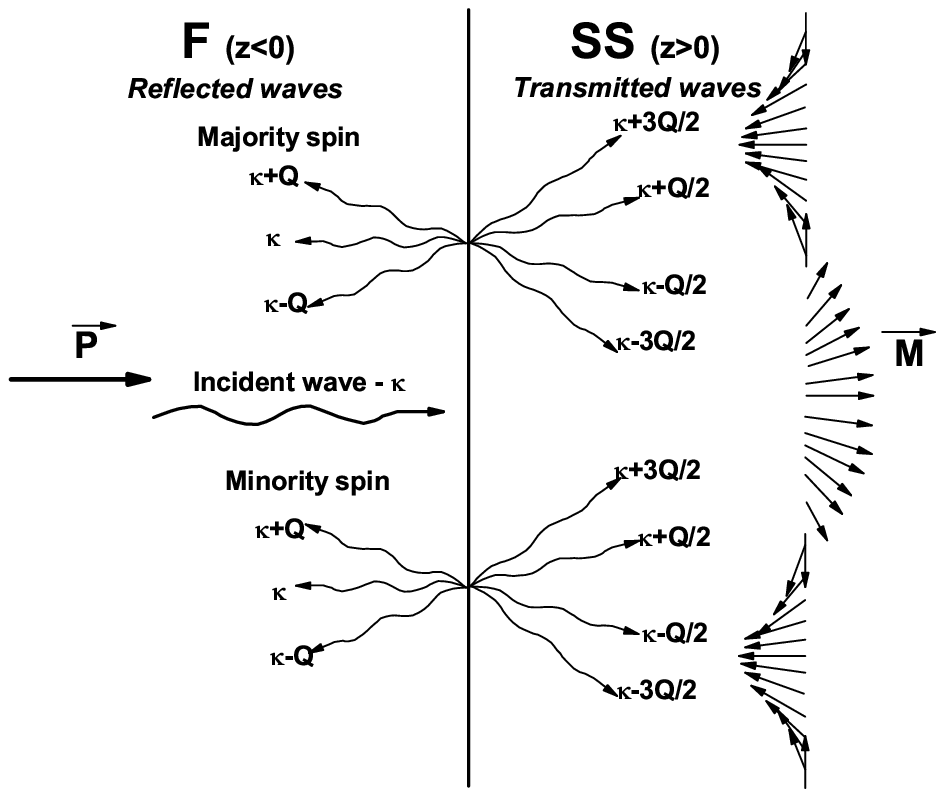}
	\caption{Cartoon of the bilayered structure. The left semi-infinite layer is a ferromagnet with a homogeneous magnetization and the right semi-infinite layer is a spin spiral with wavelength 2$\pi Q^{-1}$. The spin-polarized electrons undergo spin-diffraction at the interface.}\label{fig:fig1}
\end{figure}

In this case, the adiabatic approximation is no more valid because an electron moving through the interface keeps the memory of its spin state for some distance. Spin-polarized electrons moving from F into SS undergo "spin diffraction" as represented on Fig. \ref{fig:fig1}: an impinging electron with an in-plane incident wavevector $\kappa$ gives rise to transmitted (reflected) waves with in-plane wavevectors $\kappa+nQ/2$ ($\kappa+nQ$), 2$\pi$Q$^{-1}$ being the wavelength of the spin spiral. From the interference of all these waves, one can expect original electrical properties like non-zero torque acting on the magnetization of SS, local longitudinal spin current and even charge current (Hall effect and spin Hall effect \cite{hall}) along the F/SS interface.\par

To model the spin-dependent transport, we use Keldysh out-of-equilibrium technique\cite{manchon,keldysh} which expresses the lesser Keldysh Green functions $G^{-+}_{\sigma\sigma'}(\textbf{r}\textbf{r'})$ as a function of the basis of wavefunctions $\Psi_{l,r\sigma}(\textbf{r})$ for an electron moving from the left (right) to the right (left) reservoir:
\begin{equation}
G^{-+}_{\sigma\sigma'}(\textbf{r}\textbf{r'})=f_l(\mu_l)\Psi^*_{l\sigma'}(\textbf{r'})\Psi_{l\sigma}(\textbf{r})+f_r(\mu_r)\Psi^*_{r\sigma'}(\textbf{r'})\Psi_{r\sigma}(\textbf{r})
\end{equation}
where $f_{l(r)}(\mu_{l(r)})$ are the Fermi distribution functions in the left and right electrodes and $\mu_{l(r)}$ are the chemical potentials in these electrodes so that $V=(\mu_l-\mu_r)/e$. The electrical current density $\textbf{J}^e$, spin density ${\bf m}$ and torque ${\bf T}$ (exerted on SS) are given by the usual local definitions:
\begin{eqnarray}
&&\textbf{J}^e=\Im[(\nabla_\textbf{r'}-\nabla_\textbf{r})(G^{-+}_{\uparrow\uparrow}(\textbf{r}\textbf{r'})+G^{-+}_{\downarrow\downarrow}(\textbf{r}\textbf{r'}))]\\
&&m_x+im_y=<\sigma^+>=2G^{-+}_{\uparrow\downarrow}(\textbf{r}\textbf{r'})\\
&&m_z=<\sigma^z>=G^{-+}_{\uparrow\uparrow}(\textbf{r}\textbf{r'})-G^{-+}_{\downarrow\downarrow}(\textbf{r}\textbf{r'})\\
&&{\bf T} =\frac{\gamma J_{sd}}{\hbar \mu_B}{\bf M}\times {\bf m}
\end{eqnarray}
where $\gamma$ is the gyromagnetic ratio, $\mu_B$ is the Bohr magneton and $J_{sd}$ is the $s-d$ exchange coupling. The torque ${\bf T}$ possesses two components: the usual spin transfer torque term (STT, adiabatic torque propotionnal to $m_y$ and lying in the ($x,z$) plane), and the field-like term (IEC, non adiabatic torque perpendicular to the ($x,z$) plane). The Hamiltonian of the system is then:
\begin{equation}
	H=\frac{p^2}{2m}+U+J_{sd}(\bm{\sigma}.{\bf S})\label{e:20}
\end{equation}
where $U$ is the potential profile, $\bm{\sigma}$ is the vector of Pauli matrices and ${\bf S}$ is the magnetization of the layer (${\bf S}={\bf M}$ in SS and ${\bf S}={\bf P}$ in F). The wavefunctions are the solutions of the Schrödinger $H\Psi=E\Psi$. To solve these equation, we used the procedure developped by Calvo\cite{calvo} in a spin spiral. The boundary conditions of these wavefunctions and their derivatives lead to recurrent relations between the coefficients of these wavefunctions \cite{af}. For example, the wavefunctions of an initially majority electron originating from the left reservoir with in-plane wavevector $\kappa$ are:
\begin{widetext}
\begin{eqnarray}\label{eq:psia}
\Psi^{\uparrow(\uparrow)}_{l SS} && =\sum_n[c_{3(\kappa+Q(n-\frac{1}{2}))}(\cos\frac{\theta(x)}{2}-i\phi_{\kappa+Q(n-\frac{1}{2})}\sin\frac{\theta(x)}{2})e^{ik_{3(\kappa+Q(n-\frac{1}{2}))}z}\nonumber\\
&& +c_{4(\kappa+Q(n-\frac{1}{2}))}(\phi_{\kappa+Q(n-\frac{1}{2})}\cos\frac{\theta(x)}{2}+i\sin\frac{\theta(x)}{2})e^{ik_{4(\kappa+Q(n-\frac{1}{2}))}z}]e^{i(qy+(\kappa+Q(n-\frac{1}{2}))x)}\\
\Psi^{\downarrow(\uparrow)}_{l SS}&&=\sum_n[c_{3(\kappa+Q(n-\frac{1}{2}))}(\sin\frac{\theta(x)}{2}+i\phi_{\kappa+Q(n-\frac{1}{2})}\cos\frac{\theta(x)}{2})e^{ik_{3(\kappa+Q(n-\frac{1}{2}))}z}\nonumber\\
&&+c_{4(\kappa+Q(n-\frac{1}{2}))}(-i\cos\frac{\theta(x)}{2}+\phi_{\kappa+Q(n-\frac{1}{2})}\sin\frac{\theta(x)}{2})e^{ik_{4(\kappa+Q(n-\frac{1}{2}))}z}]e^{i(qy+(\kappa+Q(n-\frac{1}{2}))x)}
\end{eqnarray}
\end{widetext}
where $c_{3(4)}$ are complex diffraction coefficients and $k_{3(4)}$ are wavevectors for majority (minority) spin projection in SS. This procedure will be developped in a forthcoming article \cite{af}.\par

For the numerical simulations, we took parameters corresponding to spin transport in Co: the Fermi wavevectors for majority and minority spins are respectively $k_F^{\uparrow}=1.1$ \AA$^{-1}$, $k_F^{\downarrow}=0.6$ \AA$^{-1}$; the inverse wavelength of the spin spiral is $Q^{-1}=(2\pi)^{-1}$ \AA$^{-1}$ (highly non homogeneous magnetic system). The Fermi wavevectors of SS and F are the same and we consider the linear approximation : for a small enough bias voltage, only the electrons originating from the left reservoir with an energy located between $\mu_l$ and $\mu_r$ significantly contribute to the charge and spin transport. We set $\mu_l-\mu_r=38$ meV.\par

\begin{figure}[h]
	\centering
		\includegraphics[width=8cm]{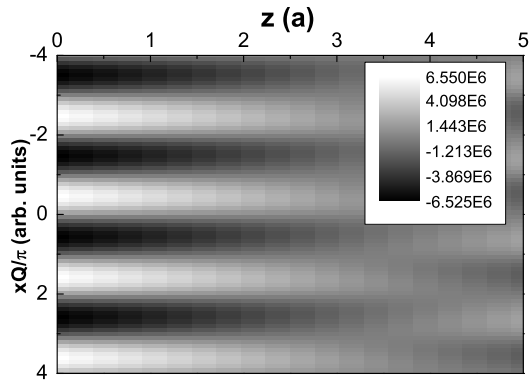}
\caption{Map of the longitudinal electrical current density in the ($x,z$) plane.}\label{fig:Jxmap}
\end{figure}

Figure \ref{fig:Jxmap} displays the longitudinal charge current $J_x^e$ in the ($x,z$) plane. Close to the interface, $J_x^e$ oscillates with SS magnetization, but rapidly decreases and vanishes to zero within 5 \AA. We observe the same behaviour for the perpendicular current $J_z^e$ except it reaches an averaged value within 5 \AA. We observe the same features for STT and IEC (not shown here).\par

This rapid decay is due to interferences between diffracted electron waves. Close to the interface, the averaging effect is smaller than in the bulk SS, so the oscillation amplitude of $J^e_x(z)$ and STT, IEC is always higher near the interface than in the bulk. \par
\begin{figure}[h]
 \centering
		\includegraphics[width=8cm]{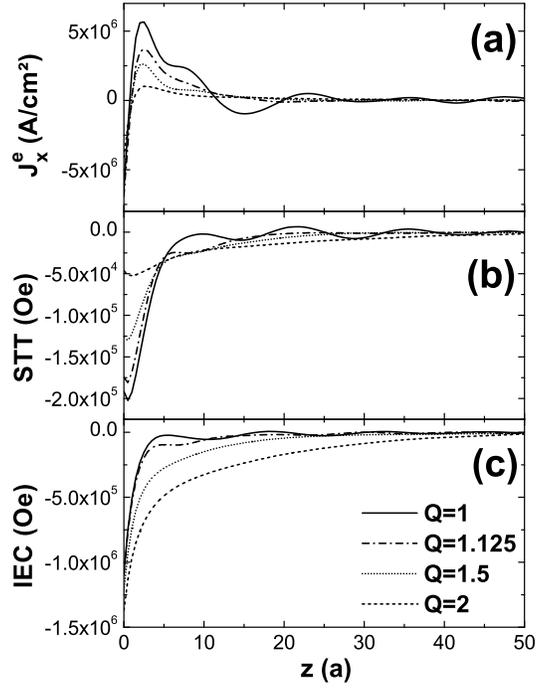}
	\caption{Longitudinal current (a), usual spin transfer torque (b) and current-induced interlayer exchange coupling (c) as a function of $z$ for different Q (see inset) and calculated at $x=\pi/2Q$.}\label{fig:fig3}
\end{figure}

These damped oscillations strongly depend on the SS wavelength 2$\pi Q^{-1}$\cite{af}. Fig. \ref{fig:fig3} displays the $z$-dependence of the longitudinal current $J^e_x$ [Fig. \ref{fig:fig3}(a)], STT [Fig. \ref{fig:fig3}(b)] and IEC [Fig. \ref{fig:fig3}(c)] for different values of SS wavelenght Q$^{-1}$.

When Q increases, the oscillation amplitude of $J^e_x$ is reduced so that $J^e_x$ vanishes more rapidly to zero: the increase of the non adiabaticity induces a more important averaging effect due to interference between multiple diffracted waves. STT and IEC have an opposite behaviour when varying Q. The oscillation amplitude of STT decreases when Q increases (similarly to $J^e_x$), whereas the amplitude of IEC increases. This illustrates the different nature of STT and IEC: STT is the adiabatic torque (vanishes in highly non homogeneous structure) and IEC is the non adiabatic torque (vanishes in adiabatic systems).\par
This study demonstrates that spin diffraction gives rise to complex characteristic in spin torque and electrical currents and is of seminal importance in non adiabatic magnetic systems.\par

This work was partially supported within the European MRTN SPINSWITCH CT-2006-035327 and the Russian Fundings for Basic Research 07-02-00918-a.

\clearpage
\clearpage
\begin{center}
\textbf{FIGURE CAPTIONS}
\end{center}

\end{document}